# DEVELOPING e-LEARNING MATERIALS FOR SOFTWARE DEVELOPMENT COURSE

Hao Shi

School of Engineering and Science, Victoria University, Melbourne, Australia
`hao.shi@vu.edu.au`

## ABSTRACT

*Software Development is a core second-year course currently offered to undergraduate students at Victoria University at its five local and international campuses. The project aims to redesign the existing course curriculum to support student-centred teaching and learning. It is intended to provide a learning context in which learners can reflect on new material, discuss their tentative understandings with others, actively search for new information, develop skills in communication and collaboration, and build conceptual connections to their existing knowledge base. The key feature of the cross-campus curriculum innovation is the use of Blackboard, short for Blackboard Learning System, to assist in course content organization and online delivery. A well-defined and integrated case study is used throughout the course to provide realistic practical experience of software development. It allows students to take control of their own learning while at the same time providing support to those students who have particular learning difficulties. In this paper, the developed curriculum and the learning outcome are described. The e-Learning material and various Blackboard tools used for teaching and learning activities are presented. Finally, conclusion is drawn from classroom experience.*

## 1. INTRODUCTION

In the past, the format of that course followed the information transmission mode, where the lecturer developed a set of lecture notes some of which were available for downloading from a web-page in essentially a one-way transmission from the lecturer to the student. The course was premised on the notion that information equates to knowledge and that providing information equates to teaching. No real consideration was given to different learning modalities of students or to the provision of materials to support students who experienced difficulties with their studies.

Student-centred teaching methods shift the focus of activity from the teacher to the learners [1]. It has changed "one size fits all" approach that offers "equal" services to all students regardless their learning abilities and skills [2]. This paradigm shift has been taking place in higher education over the last two decades [3] associated with adoption of student-centred teaching model based on the constructivist theory of learning [4].

The student-centred has made a wide implication in all the disciplines since it is increasingly being encouraged in higher education. For example, in Mathematics education, a student-assisted teaching, or peer teaching becomes common in first year mathematics courses where advanced students are recruited to be peer teachers or peer tutors for less advanced students [5]. Outcomes of student-assisted teaching are that the students learn, retain more and are more satisfied with their education [6]. The study on student-centred learning in an Advanced Social Work Practice Course through mixed methods [7] suggests that development of advanced practice skills enhances student learning and students appreciates the cooperative learning approach, finds assessment meaningful, and learning authentic. A student-centred approach is used in the MBA (Master of Business Administration) leadership classroom [8], which has successfully demonstrated that students are easy to develop new leadership behaviour [9]. In Information Systems education, a twelve-step template-based approach to developing student-centred teaching and assessment strategies [3] has been approved to be practically useful. Authentic assessment allows students to demonstrate skills and competencies that realistically represent problems and situations likely in their daily work life. All the above approaches to the construction of educational activities provide greater student learning and a more authentic assessment.

It is no doubt that student-centred teaching and learning leads to increased learning [9] although approaches in different disciplines towards the student-centred teaching and learning are various. The student-centred methods have repeatedly been shown to be superior to the traditional teacher-centred approach [10]. In this paper, it aims to design the eLearning materials in order to make it student-centred learning, and to place responsibility for the control of learning in the hands of the student. A well-defined and integrated case study is used throughout the course to provide realistic practical experience of software development in the real world.

## 2. BACKGROUND

Software Development is a core 2nd year course currently offered to both undergraduate and, in a modified form, postgraduate students at local, offshore, and international campuses within Victoria University. The proposed project aims to redesign the curriculum in order to make it student-centred learning, and to place responsibility for the control of learning in the hands of the student. The redesign represents a transition from a passive model of learning to a constructivist model; one that involves the active participation of the student in the learning process. Apart from the students' ability to control both the pace and the pathways of learning, the proposed model involves the student in active learning and collaboration. The encouragement of active and collaborative learning, and the development of a learning environment that is sensitive to learner needs, and that links all these elements in a coherent, meaningful, and helpful way for the students.

At the basis of this approach is a deep-seated respect for the integrity and ability of the student, and the development of skills and values that are relevant to the work place. In addition, because the approach encourages students to take responsibility for their own learning, students are better prepared for a future in which lifelong-learning is the norm.

The key feature of the cross-campus curriculum innovation is the use of Blackboard in mixed mode to assist in course content organization and online delivery. Discussion lists and email supplement group work and knowledge discovery. It is intended to carry out an effective and significant development to transformation of the learning environment within a short period.

## 3. CURRICULUM DEVELOPMENT

The curriculum of Software Development is developed according to the key elements of a curriculum: aims, teaching methods, assessment processes, and learning outcomes [11 and 12] as shown in Figure 1. The objectives of the course are to encourage students to use the Internet, understand the user requirements for software development, and appreciate the changing concepts of software design. The adopted approach can be described as a top-down approach because the intent of the planned change is to re-conceptualize the whole design of the curriculum.

The learning outcomes of the course can be summarized as empowering students with the ability:

- ✓ to apply an Object Oriented development methodology
- ✓ to design software applications that involve relational database and a high level processing language;
- ✓ to understand and implement rapid application development concepts and technologies;
- ✓ to develop efficiency and effectiveness in problem-solving;
- ✓ to work collaboratively and cooperatively in a team environment;
- ✓ to articulate a vision for learning and manage their own learning process;
- ✓ to develop appropriate professional and ethical judgments in the conduct of their work;
- ✓ to appreciate social and cultural diversity.

The teaching methods and assessment are designed to support the achievement of these learning outcomes. In particular, the redesign of the curriculum has incorporated an approach to teaching and learning that explores theoretical concepts in the context of the solution of a real world problem.

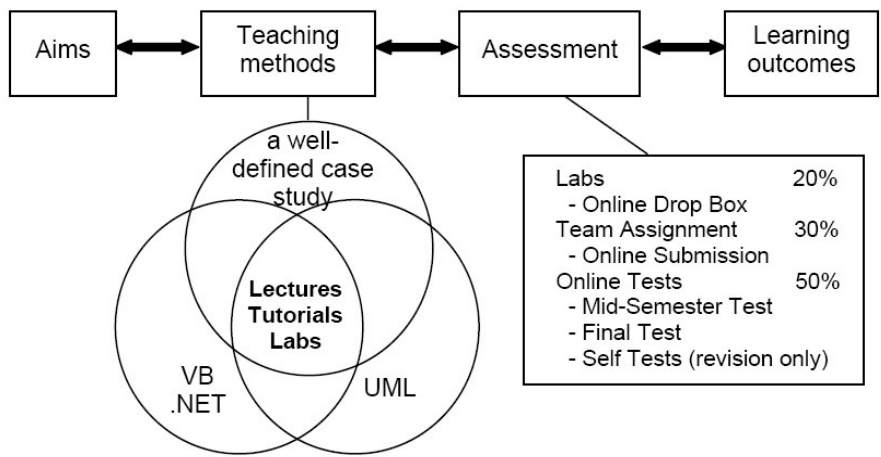

Figure 1. The curriculum of Software Development

## 4. DEVELOPING E-LEARNING MATERIALS

The key features of the cross-campus curriculum innovation is the use of Blackboard to facilitate course content organization and delivery in order to assist students in understanding the rapid development technologies/concepts for software design using a state of the art programming language. Blackboard is used in mixed mode to complement face-to-face teaching and to encourage students to use the Internet and understand the user requirements for software development. It enables the provision of links and facilities that aid the learning of national and international students with diverse backgrounds.

Blackboard is also used to integrate the learning activities of the course. Blackboard tools, such as Syllabus, Course Content, Calendar, Self Test, Quiz and Assignments, are used to support different e-Learning activities as listed in Table 1.

Table 1 Blackboard tools and student activities

| Tool | How this tool is used | Student Learning activities |
|---|---|---|
| Syllabus 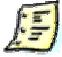 | It contains:<br>• lecturer information<br>• course aims and objectives<br>• course requirements<br>• week-by-week program<br>• textbooks<br>• assessment & deadlines<br>• course policies | Students access the course information and have a clear view of the course aim and objectives, milestones, weekly program, assessment and deadlines. Lecturer can schedule events interactively with the students. |
| Content module 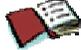 | Lectures and tutorials are delivered face-to-face but the teaching materials are available online. | Students are able to access the course material online anytime and anywhere. It facilitates diverse talents and ways of learning. |
| Calendar 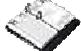 | This tool supports a global study plan for the course, course resources and external URLs. | It helps students to plan their study, meet the assessment deadlines and finds the external online resources. |
| Self Test 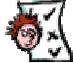 | Self-test questions are available for each learning module, which test the knowledge and learning objectives for that module and provide immediate feedback. The practice questions help students to prepare for the quiz. | It allows students to do self-evaluation, self-review, and receive instant feedback. It also provides timely feedback to the lecturer on student learning. |
| Quiz 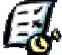 | Quiz tool is used for formal assessment at the end of each learning module. | Quiz is scheduled during the laboratory hour. Quiz uses multiple choice format so that the questions are automatically marked and the feedback available to the student immediately. |
| Drop Box 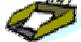 | The assignments and final project are submitted via Blackboard at regular intervals over the semester. | It allows the lecturer to distribute course assignments/project to students and for students to submit their responses electronically. |

The outcome can be summarized as follow:

(1) the development of a curriculum for SCM2313 Software Development that
- ✓ utilizes the properties of Blackboard to deliver, monitor, foster, and enhance the learning experiences of students;
- ✓ contains a syllabus and set of online materials that is consistent and uniform across all delivered campuses;
- ✓ incorporates a collection of learning materials that form the basis for a flexible learning alternative for both on-shore and off-shore students;
- ✓ supports self-directed learning and
- ✓ to encourages students to take an active approach to their learning;
- ✓ fosters the development of team skills through cooperative and collaborative learning;
- ✓ provides exposure to real world problems and engagement with real world solutions.

(2) a template for the curriculum development of a computing subject in flexible learning mode, that can be utilized by staff of the School

Blackboard is used to integrate the learning activities of the course as illustrated in Figure 2 and 3. The materials support the learning of all students and encourage students to take responsibility for their own learning. This transformation of the learning environment is made feasible through the advent of technology like Blackboard, but it provides the impetus for a complete restructure of the learning environment, and in the process address issues of resources and educational responsibility that are crucial to the ongoing development of Victoria University.

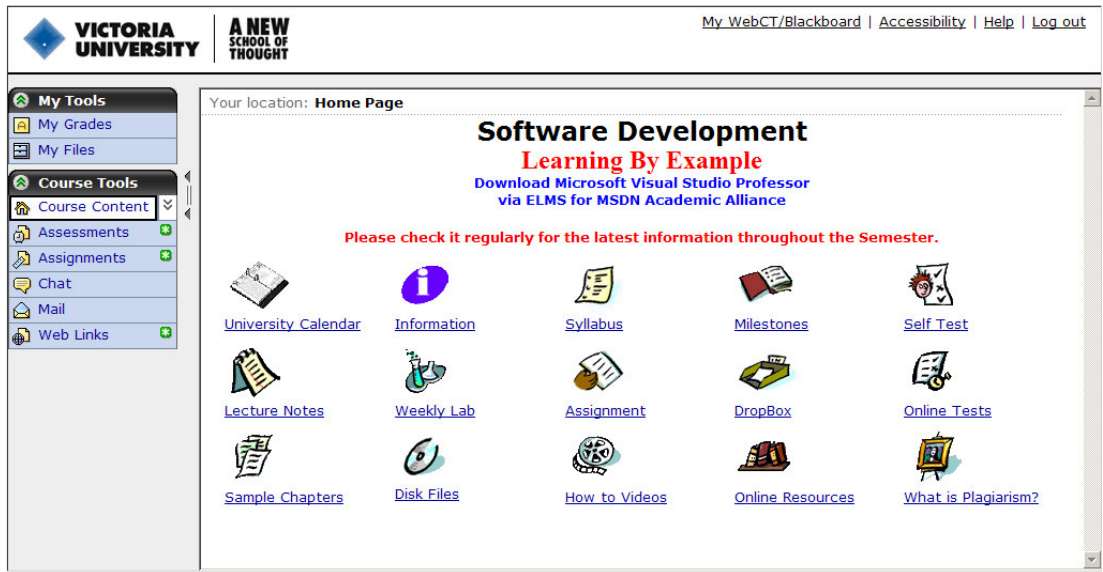

Figure 2. The homepage of Software Development Course

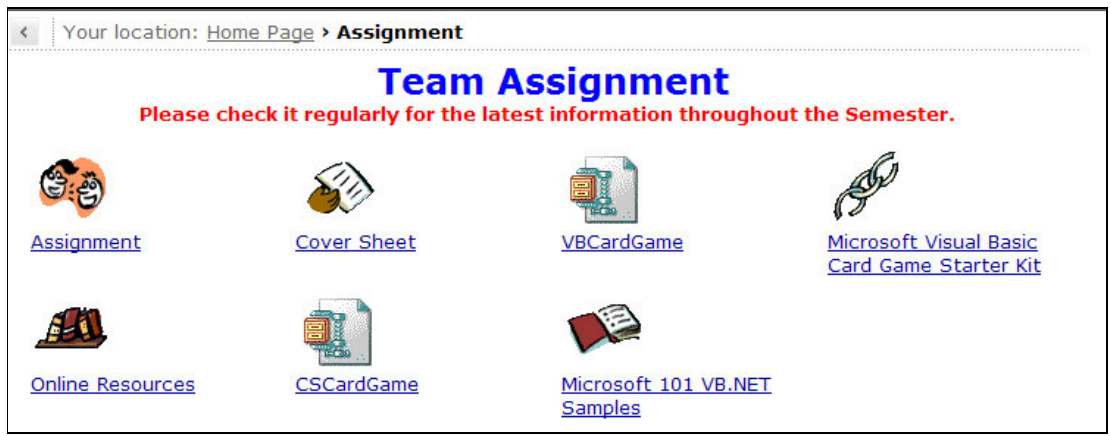

Figure 3. The Assignment Page of Software Development Course

## 5. CONCLUSIONS

The project is an effective and significant development because it proposes the redesign of the curriculum in order to make it student-centred learning, and to place responsibility for the control of learning in the hands of the student. The redesign represents a transition from a passive model of learning to a constructivist model. The proposed model encourages active and collaborative learning and the development of a learning environment that is sensitive to learner needs. Discussion lists and email supplement group work and knowledge discovery. The project provides consistent and uniform learning materials across all the delivered campuses. The e-Learning materials produced during the completion of the project form the basis of a flexible learning alternative for both on-shore and off-shore students. Its outcomes are applicable to other courses in the School. The project provides a template for the curriculum development of a computing course in flexible learning mode that can be utilized by staff of the School and the University.

### ACKNOWLEDGEMENTS

The author would like to thank Dr. John Horwood for his support throughout implementation of the project. This project is supported by the Curriculum Innovation Grant of Victoria University.

**Authors**

Dr. Hao Shi is an Associate Professor in the School of Engineering and Science at Victoria University, Australia. She completed her PhD in the area of Computer Engineering at the University of Wollongong and obtained her Bachelor of Engineering degree from Shanghai Jiao Tong University, China. She has been actively engaged in R&D and external consultancy activities. Her research interests include p2p Networks, Location-Based Services, Web Services, Computer/Robotics Vision, Visual Communications, Internet and Multimedia Technologies.

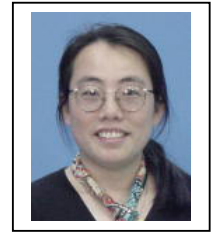